\documentclass[times,5p,twocolumn]{elsarticle}
\usepackage{amsmath}
\usepackage{graphicx}

\journal{Physics Letters B}
\newcommand{\be}{\begin{equation}}
\newcommand{\ee}{\end{equation}}
\newcommand{\bearray}{\begin{eqnarray}}
\newcommand{\eearray}{\end{eqnarray}}
\newcommand{\bse}{\begin{subequations}}
\newcommand{\ese}{\end{subequations}}
 
\begin{document}

\title{Positronium energy levels at order $m \alpha^7$: vacuum polarization corrections in the two-photon-annihilation channel}

\author{Gregory S. Adkins \fnref{email}}
\author{Christian Parsons}
\author{M. D. Salinger}
\author{Ruihan Wang}
\address{Franklin \& Marshall College, Lancaster, Pennsylvania 17604 United States}
\fntext[email]{Corresponding author e-mail address: gadkins@fandm.edu (G. Adkins)}

\date{\today}

\begin{abstract}
We have calculated all contributions to the energy levels of parapositronium at order $m \alpha^7$ coming from vacuum polarization corrections to processes involving virtual annihilation to two photons.  This work is motivated by ongoing efforts to improve the experimental determination of the positronium ground-state hyperfine splitting.
\end{abstract}

\begin{keyword}
quantum electrodynamics (QED) \sep positronium \sep hyperfine splitting (hfs)
\PACS 36.10.Dr \sep 12.20.Ds
\end{keyword}

\maketitle


\section{Introduction}
\label{introduction}

Positronium, the electron-positron bound state, is a particularly simple and interesting system.  The constituents of positronium have no known internal structure.  The properties of positronium are governed almost completely by QED--strong and weak interaction corrections are below the level of current interest due to the small value of the electron mass.  On the other hand, some features of positronium tend to complicate the analysis compared to, say, hydrogen or muonium.  The no-recoil approximation is not relevant for positronium--the mass ratio for positronium takes its maximum value of one.  Also, because positronium is composed of a particle and its antiparticle, it exhibits real and virtual annihilation into photons.  The states of positronium can be taken to be eigenstates of the discrete symmetries charge parity and spatial parity, making positronium useful in searches for new, symmetry breaking interactions.  Because of its unique properties and accessibility to high-precision experiments, positronium is an ideal system for tests of the bound state formalism in quantum field theory and for searches for new physics in the leptonic sector.

Since its discovery in 1951 \cite{Deutsch51}, positronium has been the object of increasingly precise measurements of the ground state hyperfine splitting (hfs), orthopositronium (spin-triplet) and parapositronium (spin-singlet) decay rates and branching ratios, $n=2$ fine structure, and the $2S-1S$ interval.  This progress is reviewed in Refs~\cite{Berko80,Rich81,Mills90,Rich90,Dvoeglazov93,Dobroliubov93,Karshenboim04,Rubbia04,Gninenko06} with citations to the original literature.  The most precise hfs measurements were performed by two groups in the 70s and early 80s \cite{Mills75,Mills83,Ritter84}:
\bearray
\Delta E(\rm{Brandeis}) &=& 203 \, 387.5(1.6) MHz , \nonumber \\
\Delta E(\rm{Yale}) &=& 203 \, 389.10(74) MHz .
\eearray
Both of these results are based on the observation of Zeeman mixing of ortho and para states in the presence of a static magnetic field.  More recently, a great deal of work has been done both with the Zeeman approach and with other indirect and direct methods of measurement
 \cite{Baryshevsky89,Fan96,Crivelli11,Sasaki11,Ishida12,Yamazaki12, Namba12, Cassidy12}.  A new high-precision measurement utilizing the Zeeman method with improved control of systematics was recently reported \cite{Ishida14}
\be
\Delta E(\rm{Tokyo}) = 203 \, 394.2 (1.6)_{\rm{stat}} (1.3)_{\rm{sys}} MHz .
\ee
 
Theoretical work on the positronium hfs involves calculating the energy splitting by use of bound-state methods in QED.  The principal modern approach involves the definition of an effective non-relativistic theory through matching with full QED followed by a bound-state perturbation calculation in the effective theory.  The result has the form of a perturbation series in the fine structure constant $\alpha$ augmented by powers of $\ell = \ln (1/\alpha)$.  This series has the form
\bearray
\Delta E &=& m \alpha^4 \Bigl \{ C_0 + C_1 \frac{\alpha}{\pi} + C_{21} \alpha^2 \ell + C_{20} \left ( \frac{\alpha}{\pi} \right )^2 \nonumber \\
&\hbox{}& + C_{32} \frac{\alpha^3}{\pi} \ell^2 + C_{31} \frac{\alpha^3}{\pi} \ell + C_{30} \left ( \frac{\alpha}{\pi} \right )^3 + \cdots \Big \}
\eearray
where $m$ is the electron mass.  The coefficients $C_0-C_{31}$ are known analytically as reviewed in Ref.~\cite{Adkins14a}.  The most recent result,
\be
C_{31} = -\frac{17}{3} \ln 2 + \frac{217}{90} ,
\ee
was obtained by three groups in 2000 \cite{Kniehl00,Melnikov01,Hill01}.  The numerical value of the theoretical prediction, including terms through $C_{31}$, is
\be
\Delta E(\rm{th}) = 203 \, 391.69 MHz
\ee
with an uncertainty due to uncalculated terms that has been estimated as $0.16MHz$ \cite{Melnikov01}, $0.41 MHz$ \cite{Kniehl00}, or $0.6 MHz$ \cite{Adkins14a}.  Theory and the older experiments are separated by $2.6 \sigma$ and $3.5 \sigma$ in terms of the experimental uncertainties, but theory and the new experiment are consistent with one another.

The naive size of $O(m \alpha^7)$ corrections is only $m \alpha^4 (\alpha/\pi)^3 = 4.39 kHz$, but contributions as large as several tenths of a $MHz$ have been found coming from ``ultrasoft'' energy scales \cite{Marcu11,Baker14}.  Additional $O(m \alpha^7)$ contributions have recently been obtained \cite{Adkins14a,Eides14,Adkins14b,Eides15}.  The present work is a contribution to a systematic calculation of all corrections at $O(m \alpha^7)$ begun in anticipation of yet more precise measurements of the positronium hfs.

Contributions to the positronium hfs at $O(m \alpha^7)$ can be classified as either annihilation or exchange depending on the presence or not of virtual annihilation $e^+ e^- \rightarrow n \gamma \rightarrow e^+ e^-$ in the description of the process.  Among annihilation contributions, ones that involve virtual annihilation to an odd number of photons only affect orthopositronium according to charge conjugation symmetry, while ones involving an even number of photons only affect parapositronium.  We consider here two-photon-annihilation processes affecting parapositronium and focus specifically on processes containing a vacuum polarization correction to one or both of the annihilation photons.  This set of contributions forms a gauge invariant set, and furthermore is insensitive to the particular bound-state formalism used in its evaluation, and consequently it comprises a reasonable set of contributions to be evaluated in isolation from other types of terms.  Most of the two-, three-, and four-photon annihilation contributions have non-vanishing imaginary parts, as can be seen from Cutkosky analysis \cite{Cutkosky60}.  However, the vacuum polarization function vanishes for small $k^2$ (where $k^\mu$ is the photon momentum), so the vacuum polarization corrections discussed here vanish for annihilation to on-shell virtual photons.  These contributions to the energy shift $\Delta E$ are purely real and that fact simplifies their evaluation considerably.


\section{Pure vacuum polarization corrections}
\label{pure_vp}

\begin{figure}
\includegraphics[width=8.6cm]{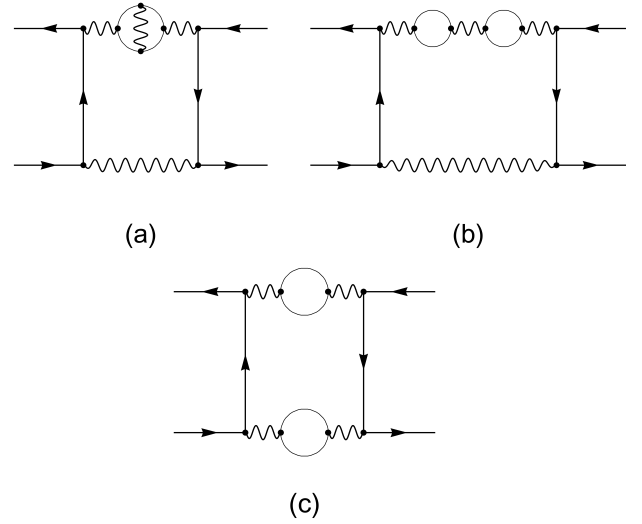}
\caption{\label{fig1} The three types of $O(m \alpha^7)$ pure vacuum polarization corrections in the two-photon-annihilation channel.  Graph (a) represents the contribution of the one-photon-irreducible two-loop vacuum polarization function.  This function is represented by a single diagram, but there are two additional contributions that can be described as self-energy corrections inside the vacuum polarization loop that are not shown.  Graph (b) gives the one-photon-reducible contribution.  Graph (c) shows a final two-loop vacuum polarization correction, one with a one-loop correction on each virtual photon.  Graphs with crossed photons are not displayed--the crossed photon graphs give energy contributions equal to those of the graphs that are displayed.  In addition, the contributions of graphs (a) and (b) must be doubled since the correction could occur on either virtual photon.}
\end{figure}

The ``pure vacuum polarization'' corrections involving either the two-loop vacuum polarization function or a product of two one-loop functions are shown in Fig.~\ref{fig1}.  
The effect of a vacuum polarization (VP) correction is to modify a photon propagator according to
\be
\frac{1}{p^2} \rightarrow \frac{1}{1+\Pi_R(p^2)} \frac{1}{p^2} = \frac{1}{p^2} - \Pi_R(p^2) \frac{1}{p^2} + \cdots
\ee
where the renormalized scalar vacuum polarization function can be expressed in a spectral form as
\be \label{def_spectral_form}
\Pi_R(p^2) = \int_0^1 dv g(v) \frac{p^2}{p^2-4m^2/(1-v^2)} .
\ee
At one-loop order the spectral function $g(v)$ takes the form
\be \label{one_loop_vp}
g_1(v) = - \frac{v^2 (1-v^2/3)}{1-v^2} \frac{\alpha}{\pi} .
\ee
The two-loop function has been given by K\"all\'en and Sabry \cite{Kallen55}, Schwinger \cite{Schwinger73}, and in a form using ``standard'' notation for the dilogarithm function \cite{Lewin81}, by Eides, Grotch, and Shelyuto \cite{Eides01}.  One must note that the K\"all\'en-Sabry form is for the reducible vacuum polarization function (Figs.~1a plus 1b), not the irreducible function of Fig.~1a alone.  The Schwinger and Eides-Grotch-Shelyuto forms are for the irreducible function.

The energy shift due to the vacuum polarization graphs considered in this article is insensitive to the particular bound state formalism we employ because the energy and momentum values that contribute are purely ``hard''--of order $m$, not $m \alpha$ or $m \alpha^2$.  We choose to use the formalism of Ref.~\cite{Adkins99}, in which the energy shift is an expectation value
\be \Delta E = i \bar{\Psi} \delta K \Psi .
\ee
The two-body states $\Psi$, $\bar{\Psi}$ carry information about the positronium spin state $\chi$:
\be
\Psi \rightarrow \phi_0 \begin{pmatrix} 0 & \chi \\ 0 & 0 \end{pmatrix}
\quad \bar \Psi^T \rightarrow \phi_0 \begin{pmatrix} 0 & 0 \\ \chi^\dagger & 0 \end{pmatrix} ,
\ee
where $\chi$ is the two-by-two two-particle singlet spin matrix $\chi = 1/\sqrt{2}$.  The positronium states, in this approximation, have vanishing relative momentum, and $\phi_0=\sqrt{m^3 \alpha^3/(8 \pi n^3)}$ is the wave function at spatial contact for a state of principal quantum number $n$ and orbital angular momentum $\ell=0$.

The explicit expression for the energy shift for the vacuum polarization diagram of Fig.~\ref{fig1}a is
\bearray \label{energy_1a_explicit}
\Delta E_{1a} &=& (-1) 4 i \phi_0^2 \int \frac{d^4 p}{(2 \pi)^4} \bigl ( -\Pi_R(p^2) \bigr ) \frac{-i}{p^2} \frac{-i}{(P-p)^2} \nonumber \\
&\times& \mathrm{tr} \Bigl [ \begin{pmatrix} 0 & 0 \\ \chi^\dagger & 0 \end{pmatrix} (-i e \gamma^\mu) \frac{i}{\gamma (P/2-p)-m} (-i e \gamma^\nu)   \Bigr ] \cr
&\times& \mathrm{tr} \Bigl [ (-i e \gamma^\nu) \frac{i}{\gamma (P/2-p)-m} (-i e \gamma^\mu) \begin{pmatrix} 0 & \chi \\ 0 & 0 \end{pmatrix} \Bigr ]
\eearray
where $P=(2m,\vec 0 \,)$ is the positronium 4-momentum in the center-of-mass frame, $p$ is the 4-momentum of the vacuum polarization corrected photon, the initial $(-1)$ is a fermonic minus sign, and the factor of $4$ accounts for the graph with crossed photons and for the fact that the vacuum polarization correction could act on either photon.  After evaluating the traces and some simplifications, the energy shift above takes the form
\be \label{energy_1a}
\Delta E_{1a} = \frac{m \alpha^5}{\pi} \int \frac{d^4 p}{i \pi^2} \frac{2 \vec p\, ^2 \Pi_R(p^2)}{p^2 (p-P)^2 (p^2-p \cdot P)^2}
\ee
for the ground state ($n=1$).  The integral over $p$ can be evaluated either using Feynman parameters or by the method of poles--closing the $p^0$ contour with a half-circle of infinite radius in the upper or lower half plane.  In either case, the result for the energy shift becomes
\bearray \label{1d_integral}
\Delta E &=&  \int_0^1 dv \frac{g(v)}{2v(1-v^2)} \Bigl \{ (2-v)(1+v)^2 \ln \Bigl ( \frac{1+v}{2} \Bigr ) \nonumber \\
&\hbox{}& - (2+v)(1-v)^2 \ln \Bigl ( \frac{1-v}{2} \Bigr ) \nonumber \\ 
&\hbox{}& +  2v^3 \ln v + 2v (1-v^2) \Bigr \} \frac{m \alpha^5}{\pi}.
\eearray
For the one-loop vacuum polarization of (\ref{one_loop_vp}) the energy shift is
\be \label{IVP}
\Delta E_{VP;1} = I_{VP} \frac{m \alpha^6}{\pi^2}, \quad I_{VP} = -\frac{1}{6} \zeta(2)
\ee
in accord with the result of \cite{Adkins93}.  The two-loop irreducible vacuum polarization contribution of Fig.~1a is
\bearray \label{PSLQ_for_1a}
\Delta E_{1a} &=&  \Bigl \{ -\frac{65}{24} \zeta(4) + \frac{10}{3} \zeta(2) \ln^2 2 - \frac{1}{18} \ln^4 2  \nonumber \\
&\hbox{}&  - \frac{4}{3} a_4 + \frac{161}{96} \zeta(3) - \frac{39}{8} \zeta(2) \ln 2 \nonumber \\
&\hbox{}& + \frac{475}{192} \zeta(2) - \frac{43}{96} \Bigr \} \frac{m \alpha^7}{\pi^3}
\eearray
where $\zeta$ is the Riemann zeta function and $a_4 \equiv {\rm Li}_4(1/2)$.  We didn't actually ``do'' the integral for $\Delta E_{1a}$.  Rather, we obtained a numerical result to high precision (100 digits) and used the PSLQ algorithm \cite{Ferguson99} to obtain (\ref{PSLQ_for_1a}).

For the reducible contribution of Fig.~1b we were able to actually perform the exact integral.  We used the spectral function
\be
g_{1b}(v) = \frac{2v^2(1-v^2/3)}{1-v^2} \Bigl \{ \frac{8}{9} - \frac{v^2}{3}
 - \frac{v}{2} \Bigl ( 1-\frac{v^2}{3} \Bigr ) \ln \Bigl ( \frac{1+v}{1-v} \Bigr ) \Bigr \} \frac{\alpha^2}{\pi^2}
\ee
extracted from the work of K\"all\'en and Sabry \cite{Kallen55} in (\ref{1d_integral}) to obtain
\be
\Delta E_{1b} = \Bigl \{ \frac{1}{6} \zeta(3)-\frac{2}{3} \zeta(2) \ln 2 + \frac{19}{36} \zeta(2) - \frac{41}{108} \Bigr \} \frac{m \alpha^7}{\pi^3} .
\ee

Finally, for the term of Fig.~1c with a one-loop vacuum polarization correction on each annihilation photon, we were not able to achieve an exact result.  The numerical value of this contribution is
\be
\Delta E_{1c} = -0.045140511436 \, \frac{m \alpha^7}{\pi^3} .
\ee


\section{Terms involving one-loop vacuum polarization}
\label{two_loop}

Additional vacuum polarization corrections are shown in Fig.~2.  They contain a one-loop vacuum polarization part combined with self-energy (SE), vertex, or ladder corrections to the bare two-photon-annihilation process.  These contributions form a set that is best calculated together because, after renormalization of the self-energy and vertex parts, each of these graphs contains an infrared divergence that only cancels when all four are summed.

\begin{figure}
\includegraphics[width=8.6cm]{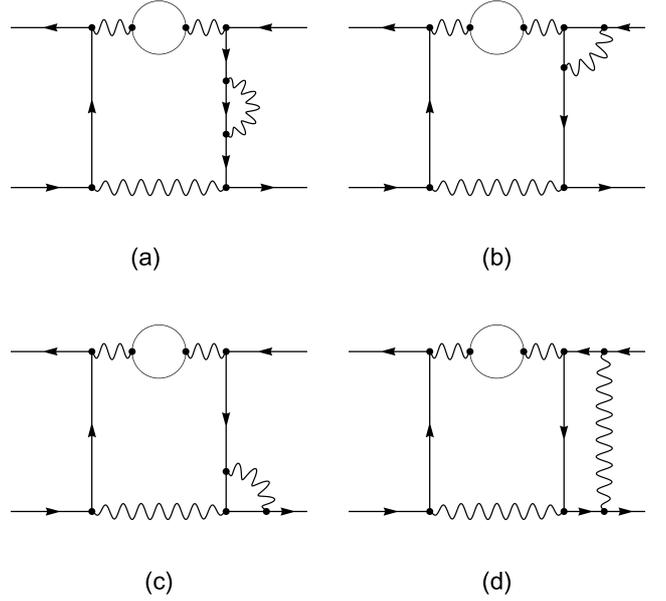}
\caption{\label{fig2} The four types of two-loop corrections to the two-photon-annihilation channel containing a one-loop vacuum polarization function.  Graph (a) contains a self-energy correction.  Graph (b) displays a correction having a vertex correction on the same photon that is modified by vacuum polarization (this is ``Vertex Type A'').  Graph (c) shows a correction where the vertex and vacuum polarization affect different virtual photons (this is ``Vertex Type B'').  Graph (d) contains a ladder correction.  Each of these graphs must be multiplied by a factor of eight to account for equivalent graphs involving crossed annihilation photons, VP corrections on either photon, and the self-energy, vertex, or ladder corrections appearing at various positions in the diagram.}
\end{figure}

Our calculational method is to start with a form like (\ref{energy_1a_explicit}) containing the one-loop vacuum polarization correction of (\ref{def_spectral_form}) and (\ref{one_loop_vp}) and tack on a self-energy, vertex, or ladder correction.  Convenient forms for the one-loop self-energy and vertex corrections were given in \cite{Adkins01}.  The renormalized self-energy function corrects the bare electron propagator according to
\be \label{se_correction}
\frac{1}{\gamma p-m} \rightarrow \frac{\alpha}{\pi} \Bigl \{ S_1 + S_2(p) \Bigr \} \frac{1}{\gamma p-m}
\ee
where
\bse
\bearray
S_1 &=& \ln \biggl ( \frac{\lambda}{m} \biggr ) + \frac{1}{2} , \\
S_2(p) &=& \int dx du f_{SE} \frac{N(p)}{D(p)}
\eearray
\ese
with $f_{SE} = -1/(2u)$ and
\bse
\bearray
N(p) &=& \left \{ 2m - (1-x) \gamma p \right \} (\gamma p+m) \\
D(p) &=& p^2-m^2-\frac{xm^2}{(1-x)u} .
\eearray
\ese
(All parametric integrals in the self-energy and vertex corrections run from $0$ to $1$.)  A photon mass $\lambda$ was introduced as an infrared regulator in the course of on-shell renormalization.  The vertex correction has the form
\be \label{vertex_correction}
\gamma^\mu \rightarrow \frac{\alpha}{\pi} \Bigl \{ V_1^\mu + V_2^\mu (p',p) + V_3^\mu (p',p) \Bigr \}
\ee
where $p$ and $p'$ are the incoming and outgoing electron momenta and
\bse \label{vertex_parts}
\bearray
V_1^\mu &=& \gamma^\mu \, \Bigl ( - \ln \Bigl ( \frac{\lambda}{m} \Bigr ) - \frac{5}{4} \Bigr ) , \\
V_2^\mu (p',p) &=& \int dx \, du \, \frac{-N^\mu}{4 H} , \\
V_3^\mu (p',p) &=& \gamma^\mu \int dx \, du \, dz \, \frac{-x (H-x m^2)}{2 {\bar H}} .
\eearray
\ese
The parametric functions are
\bse
\bearray
H &=& (1-x) \big [ u(m^2-p'^2) + (1-u)(m^2-p^2) \bigr ] \nonumber \\
&\hbox{}& - x u (1-u) (p'-p)^2 + x m^2 , \\
{\bar H} &=&x m^2 + z (H-x m^2) , \\
N^\mu &=& \gamma^\lambda \left ( \gamma (p'+Q)+m \right ) \gamma^\mu \left ( \gamma (p+Q)+m \right ) \gamma_\lambda
\eearray
\ese
where $Q = -x (u p' + (1-u) p)$.

The self-energy contribution is pictured in Fig.~2a.  The diagram shown must be multiplied by eight to account for the two places where the SE correction could occur, the two photons that the VP could correct, and the two types of graph (with uncrossed and crossed photons).  All eight contribute equally.  There are two parts to the SE contribution: $S_1$ and $S_2(p)$ of (\ref{se_correction}).  The $S_1$ contribution is $2 S_1 I_{VP} (m \alpha^7/\pi^3)$ where $I_{VP}$ is the $O(m \alpha^6)$ VP correction of (\ref{IVP}).  For the $S_2(p)$ contribution we could preform the $d^4 p$ integral as a whole using Feynman parameters, or do $d p^0$ first by poles followed by $d^3 p \rightarrow 4 \pi p^2 dp$ numerically.  Both results are shown in Table~\ref{table1}, followed by their total.

There are two classes of vertex corrections: ones with the VP and vertex corrections affecting the same photon (type A, Fig.~2b) or affecting different photons (type B, Fig.~2c).  Each has three parts as specified in (\ref{vertex_correction}) and (\ref{vertex_parts}).  The various vertex correction integrals were performed both by poles and parameters, with results recorded in Table~\ref{table1}.

\begin{table}
\begin{center}
\caption{\label{table1} Contributions to the p-Ps energy levels coming from one-loop vacuum polarization (VP) corrections combined with self-energy (SE) (Fig.~2a), vertex (Figs.~2b and c), and ladder (Fig.~2d) corrections to the two-photon-annihilations graphs.  For type A (type B) vertex corrections, the VP and vertex corrections act on the same photon (different photons).  The separate contributions to the self-energy, etc., parts are shown along with their totals, all in units of $m \alpha^7/\pi^3$.  The infrared divergence is contained in $L\equiv \ln (\lambda/m) \, I_{VP}$. }
\begin{tabular}{ccc}
Term & Poles Result & Parameters Result \\
\hline\noalign{\smallskip}
SE: 1 &  $2L-\frac{1}{6} \zeta(2)$ & $2L-\frac{1}{6} \zeta(2)$   \\
SE: 2 &  -0.76281(16) & -0.7629639(4) \\
SE & $2L-1.03697(16)$ & $2L -1.0371196(4)$ \\
\hline
Vertex A: 1     &  $-2L+\frac{5}{12} \zeta(2)$ & $-2 L+\frac{5}{12} \zeta(2)$ \\
Vertex A: 2     &  -0.2272831(5) &  -0.2272846(14) \\
Vertex A: 3     &  0.1876296(12) & 0.1876304(19) \\
Vertex A & $-2L + 0.6457357(13)$ & $-2L + 0.6457350(24)$ \\
\hline
Vertex B: 1    &  $-2L+\frac{5}{12} \zeta(2)$ & $-2L +\frac{5}{12} \zeta(2)$ \\
Vertex B: 2    &  0.0693357(4) &  0.0693348(6) \\
Vertex B: 3    &   0.2046073(11) & 0.2046053(16) \\
Vertex B       & $-2L + 0.9593322(12)$ & $-2L  + 0.9593293(18)$ \\
\hline
Ladder: 1     & $2 L + \frac{1}{3} \zeta(2)$ &  $2L  + \frac{1}{3} \zeta(2)$ \\
Ladder: 2     &  0.3124063(6) &  0.3124057(21) \\
Ladder: 3      &  -0.5448084(11) &  -0.5448063(10) \\
Ladder    & $2L + 0.3159093(13)$ & $2L + 0.3159108(24)$ \\
\end{tabular}
\end{center}
\end{table}

The ladder correction shown in Fig.~2d gives an energy shift that looks like (\ref{energy_1a_explicit}) except that the right-hand trace of (\ref{energy_1a_explicit}) is replaced by 
\bearray \label{ladder_trace}
 \mathrm{tr} \bigl [ \; \bigr ] &\rightarrow& 2  \int \frac{d^4 q}{(2 \pi)^4} \; \mathrm{tr} \Bigl [ (-i e \gamma^\beta) \frac{i}{\gamma (-P/2+q)-m} \nonumber \\
 &\times& (-i e \gamma^\nu) \frac{i}{\gamma (P/2+q-p)-m} (-i e \gamma^\mu) \nonumber \\
 &\times& \frac{i}{\gamma (P/2+q)-m} (-i e \gamma_\beta) \begin{pmatrix} 0 & \chi \\ 0 & 0 \end{pmatrix} \Bigr ] \, \Bigl ( \frac{-i}{q^2-\lambda^2} \Bigr )
 \eearray
where the factor of $2$ comes because the ladder correction could occur on either side of the diagram.  Again an infrared divergence, here arising from the binding singularity, is regulated by including a photon mass $\lambda$.  The contribution of (\ref{ladder_trace}) can be written as
\be
-2 (4 \pi \alpha)^2 \int \frac{d^4 q}{(2 \pi)^4} \frac{N(q)}{D(q) Z(q)}
\ee
where $D(q) = ((P/2-q)^2-m^2) ((P/2+q)^2-m^2) (q^2-\lambda^2)$, $Z(q) = ((P/2+q-p)^2-m^2)$, and
\bearray
N(q) &=& \mathrm{tr} \Bigl [ \gamma^\beta \left ( \gamma (-P/2+q)+m \right ) \gamma^\nu \left ( \gamma (P/2+q-p)+m \right )  \nonumber \\
&\hbox{}& \times \; \gamma^\mu \left ( \gamma (P/2+q)+m \right ) \gamma_\beta \begin{pmatrix} 0 & \chi \\ 0 & 0 \end{pmatrix} \Bigr ] .
\eearray
We isolate the small-$q$ singularity by use of the decomposition:
\bearray \label{decomp}
\frac{N(q)}{D(q) Z(q)} &=& \frac{N(0)}{D(q) Z(0)} + \frac{N(0)}{D(q)} \Bigl ( \frac{1}{Z(q)} - \frac{1}{Z(0)} \Bigr ) \nonumber \\
&\hbox{}&  + \frac{N(q)-N(0)}{D(q) Z(q)}
\eearray
consisting of terms 1, 2, and 3, which we evaluate in turn.
The first term in (\ref{decomp}) contains the infrared singular part of the $q$ integral, but otherwise is simply proportional to $I_{VP}$ because 
\be
\frac{N(0)}{Z(0)} = -4 m^2 \mathrm{tr} \Bigl [ \gamma^\nu \frac{1}{\gamma (P/2-p)-m} \gamma^\mu \begin{pmatrix} 0 & \chi \\ 0 & 0 \end{pmatrix} \Bigr ]
\ee
is proportional to the right-hand trace of (\ref{energy_1a_explicit}).  The IR-singular binding integral is \cite{Adkins02}
\be
\int \frac{d^4 q}{i \pi^2} \frac{-m^2}{D(q)} = \frac{m \pi}{\lambda} + \ln \Bigl ( \frac{\lambda}{m} \Bigr ) - 1 + O \Bigl ( \frac{\lambda}{m} \Bigr ) .
\ee
The $m \pi/\lambda$ term represents the part of the ladder correction that comes from exchange of a Coulomb photon and so is just part of the binding that created the bound state in the first place, and so must not be included again here \cite{Caswell77}.  Or, to put it another way, the $m \pi / \lambda$ term is the piece that is removed in the matching calculation if an effective field theory formalism had been used \cite{Caswell86,Adkins02}.  So term 1 contributes $2 (\ln (\lambda/m) - 1) \, I_{VP} \, (m \alpha^7/\pi^3)$ to the energy shift.  Terms 2 and 3 of the ladder correction were constructed to be IR safe and can be evaluated directly using $\lambda \rightarrow 0$.  (It is useful to average over the directions of $\vec q$ when performing the $d^3 q$ integration as described in \cite{Caswell79}.)  The results for the various parts of the ladder correction are reported in Table~1.


\section{Results and discussion}
\label{results}

The results for all VP corrected two-photon-annihilation graphs at $O(m \alpha^7)$ are shown in Table~\ref{table2}.  We see that all infrared divergences cancel in the sum. The ground state parapositronium energy level correction due to vacuum polarization effects in the two-photon-annihilation channel is
\be \label{result}
\Delta E = -0.153095(3) \frac{m \alpha^7}{\pi^3}.
\ee
The numerical value of this correction is small due to large cancellations among the various terms.  This parapositronium shift corrects the hfs (spin-triplet minus spin-singlet) by the amount
\be
\Delta E_{\rm{hfs}}= 0.67 kHz .
\ee
Higher $S$ states with principal quantum number $n$ have corrections that are the same as (\ref{result}) except divided by the $n^3$ factor that comes from the square of the wave function at spatial contact.  States with $\ell \ne 0$ are not corrected by the terms considered here at order $O(m \alpha^7)$.  Additional contributions in the two-photon-annihilation channel involve terms having a one-loop SE, vertex, or ladder correction on each side of the diagram, and two-loop self-energy, vertex, etc., corrections on a single side of the diagram.  Calculation of these terms is in progress.

\begin{table}
\begin{center}
\caption{\label{table2} Positronium energy level corrections at $O(m \alpha^7)$ coming from vacuum polarization corrections to two-photon-annihilations graphs.  The tabulated results are contributions to $I$ where $\Delta E = (m \alpha^7/\pi^3) I$.  The poles and parameters results from Table~\ref{table1} were combined to give the results shown here.  The infrared divergence contained in $L=\ln (\lambda/m) \, I_{VP}$ cancels in the sum.}
\begin{tabular}{ccc}
Term & Diagram & Energy shift \\
\hline\noalign{\smallskip}
irreducible 2-loop VP & 1a & -0.9205630 \\
reducible 2-loop VP & 1b & -0.0712481 \\
1-loop VP on each & 1c & -0.0451405 \\
VP - SE &  2a & $2L-1.0371196(4)$ \\
VP - vertex A & 2b & $-2L + 0.6457355(12)$ \\
VP - vertex B & 2c & $-2L + 0.9593313(10)$ \\
VP - ladder & 2d & $2L + 0.3159096(12)$ \\
\hline
total & & -0.153095(3) \\
\end{tabular}
\end{center}
\end{table}


\vspace{0.4cm}
We acknowledge the support of the National Science Foundation through Grant No. PHY-1404268 and of the Franklin \& Marshall College Grants Committee through the Hackman Scholars Program.


     


\end{document}